# INITIAL OBSERVATIONS OF MICROPULSE ELONGATION OF ELECTRON BEAMS IN A SCRF ACCELERATOR*

A. H. Lumpkin[#], R. Thurman-Keup, D. Edstrom Jr., J. Ruan, and J. Santucci
Fermi National Accelerator Laboratory, Batavia, IL 60510 USA

## Abstract

Commissioning at the SCRF accelerator at the Fermilab Accelerator Science and Technology (FAST) Facility has included the implementation of a versatile bunch-length monitor located after the 4-dipole chicane bunch compressor for electron beam energies of 20-50 MeV and integrated charges in excess of 10 nC. The team has initially used a Hamamatsu C5680 synchroscan streak camera to assess the effects of space charge on the electron beam bunch lengths. An Al-coated Si screen was used to generate optical transition radiation (OTR) resulting from the beam's interaction with the screen. The chicane bypass beamline allowed the measurements of the bunch length without the compression stage at the downstream beamline location using OTR and the streak camera. We have observed electron beam bunch lengths from 5 to 16 ps (sigma) for micropulse charges of 60 pC to 800 pC, respectively. We also report a compressed sub-ps micropulse case.

## INTRODUCTION

One of the more obvious effects of space-charge forces acting within micropulses in photoinjectors and SCRF linacs is the elongation of the electron bunch compared to the initial drive laser bunch length [1,2]. During the initial 20-MeV commissioning run of the Fermilab Accelerator Science and Technology (FAST) facility [3], we took advantage of a 1.5-m drift between the photoinjector rf gun and the initial accelerating capture cavity CC2. Recently, another cavity, CC1, has been installed in this drift space so that the beam is accelerated to higher energies (~50 MeV) immediately following the photoelectric injector and gun diagnostics station [4]. We now have a direct comparison available of the observed electron beam bunch lengths for different micropulse charges and the two accelerator configurations: both with the full 1.5 m drift and with the much shorter drift following installation of the additional ~1 m of accelerating structure.

The chicane bypass beamline allowed the measurements of the bunch length without the compression stage at the downstream beamline location using OTR and the streak camera. The UV component of the drive laser had previously been characterized with a Gaussian fit sigma of 3.5-3.7 ps. However, the uncompressed electron beam was observed to elongate as expected due to space-charge forces in the 1.5-m drift from the gun to the first SCRF accelerator cavity in this initial configuration. We also report our results with the CC1 cavity installed. A preliminary ASTRA-Elegant prediction is noted. Finally, we report generation of sub-ps micropulses at FAST for the first time using the 4-dipole bunch compressor (chicane, see Fig. 1).

## EXPERIMENTAL ASPECTS

Two main aspects of the experiment are the injector as the source of the electrons in a 3-MHz pulse train and the Hamamatsu C5680 streak camera configured with the 81.25 MHz synchroscan plugin unit. The beam generates the OTR at the X121 converter screen These topics will be discussed in this section.

### The Injector Linac

The high-power electron beams for the FAST facility [3] are generated in a photoelectric injector (PEI) based on a UV drive laser and the L-band rf photocathode (PC) gun cavity. The PEI drive laser is comprised of multiple stages including a Calmar Yb fiber oscillator and amplifier, several YLF-based amplification stages, a final Northrup Grumman IR amplification stage, and two frequency-doubling stages that result in a UV component at 263 nm with a nominal 3-MHz micropulse bunch structure [5]. The UV component is transported from the laser lab through the UV transport line to the photocathode of the gun for generation of the photoelectron beams for use in the SC rf accelerator.

The low-energy section of the facility and part of the first cryomodule are schematically shown in Fig. 1. After the L-band rf PC gun, the beam is accelerated through two L-band superconducting cavities resulting in a beam energy of up to 50 MeV, though initially this was limited to 20 MeV due to the absence of CC1. We will report the bunch length elongation observed downstream for both configurations.

### The Streak Camera System

The linac streak camera consists of a Hamamatsu C5680 mainframe with S20 PC streak tube and can accommodate a vertical sweep plugin unit and either a horizontal sweep unit or blanking unit. The UV-visible input optics allow the assessment of broadband OTR. A M5675 synchroscan unit with its resonant circuit tuned to 81.25 MHz from the Master Oscillator (MO) and a M5679 horizontal sweep unit were used for these studies. The low-level rf is amplified in the camera to provide a sine wave deflection voltage for the vertical plates that results in low jitter (~1ps) of the streak camera images and allows for synchronous summing of a pulse train of OTR. The temporal resolution is about 2.0 ps (FWHM), or 0.8 ps (sigma), for NIR photons at 800 nm. When combined with the C6878 phase locked loop (PLL) delay box we can track phase effects at the ps level over several tens of minutes to hours and within the macropulse to about 200 fs. Use of the 81.25-MHz-locked synchroscan unit, used previously at the A0PI [6], has also been applied at FAST for these studies.

___
* This work was supported by the DOE contract No.DEAC02-07CH11359 to the Fermi Research Alliance LLC.
# lumpkin@fnal.gov

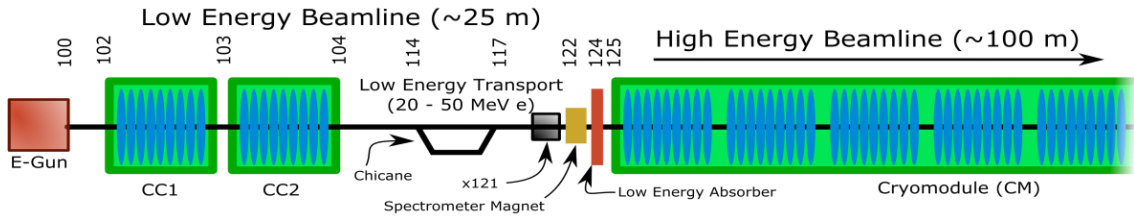

Figure 1: Schematic of the low-energy FAST electron beamline. The positions noted above the beamline correspond to instrumentation stations corresponding to the distance from the PEI gun (beamline location 100). The location of the X121 OTR station is indicated.

Locking of the synchroscan to the 81.25 MHz clock signal in this way provides essential stability to the system. As a point of comparison, the M5676 vertical fast sweep unit has ~20-ps of internal trigger jitter in addition to ~100-ps trigger jitter of a DG535 delay unit. When running on the fastest sweep range with a full scale range of 150 ps, the sum results in highly unreliable placement of the streak image within the visible frame of the readout camera. This is performed with a Prosilica 1.3-Mpix Gig-E vision digital CCD which is thus compatible with the video acquisition system [7,8] designed for all of the FAST beamline imaging stations. Commissioning of this readout camera with the laser lab streak camera, as well as the image analysis tools, was one of the goals reported previously [9].

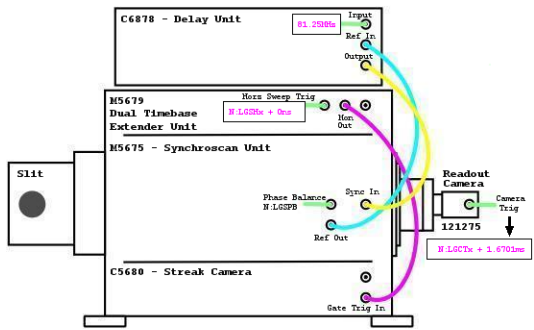

Figure 2: A streak camera wiring diagram. N:LGSHx and N:LGCTx are ACNET names for beam-synchronized VME-based timers.

A second set of deflection plates provides the orthogonal deflection for the slower time axis in the 100-ns to 10-ms regime. These plates are driven by the M5679 dual-axis sweep unit which was also commissioned during these studies. In order to assess effects on the scale of a single micropulse, we used the dual-sweep mode for separating the 3 MHz micropulses. The diagram of the final streak camera configuration in Fig. 2 is representative of all studies noted hereafter. Unless noted otherwise, the streak camera's synchroscan unit was phase locked to the master oscillator, which operationally provides the rf sync for the linac and rf gun.

Streak camera calibrations were performed with the new Prosilica readout camera through use of a Colby delay unit (PN109122; SN#8081195) to provide discrete and known ps-regime delay changes. We took 10-, 20-, and 30-ps steps and averaged the mean positions of the images. The resulting calibration factors with 1x1 pixel binning in the CCD were: R4: 1.82 ± 0.1ps/pixel, R3: 1.36 ± 0.1 ps/pixel, R2: 0.68±0.03 ps /pixel, R1: 0.38±0.03 ps/pixel. We subsequently adjusted the R1 rf amplifier gain to improve the calibration factor's measured value to be 0.10 ps/pixel for the sub-ps bunch length study only.

## EXPERIMENTAL RESULTS

### Initial Bunch lengths with Drift (no CC1)

Our studies began with the evaluation of the observed bunch lengths with the 1.5-m drift to CC2 by using the X121 OTR with transport to the linac streak camera. To evaluate the possible charge density effects at the photocathode, we tracked the variation of the 20 MeV electrons bunch length for a UV laser spot size of 260 microns in Fig. 3 and at 550 microns in Fig. 4. In Fig. 3 we show the results using the R2 streak range with the largest bunch length of 10 ps at 0.45 nC.

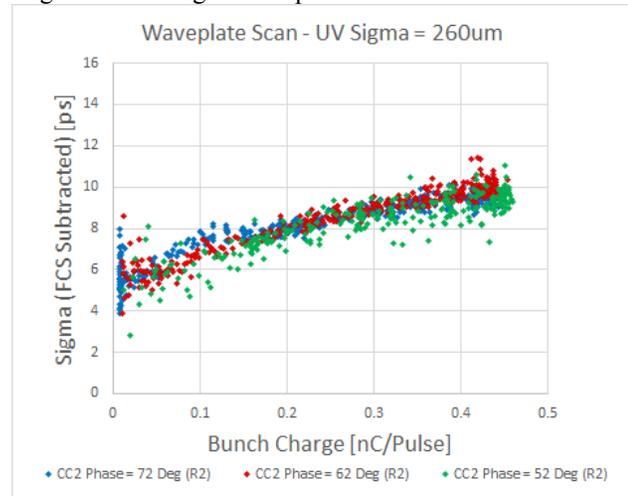

Figure 3: Example of bunch length vs charge at X121 for two streak ranges R1 and R2. The UV laser spot was 260 μm (σ).

The laser lab transmission waveplate for the UV was adjusted to vary the energy delivered to the photocathode, and hence the charge. Note more charge is generated with the larger laser spot as expected, and the off crest data at 52 deg are shorter up to 0.5 nC, even without the chicane being involved as seen in Fig. 4.

In Fig. 5, we show a direct comparison of the trends of elongation without and with CC1 installed. The effect is clearly more extreme when the 1.5-m drift condition is present, and at 800 pC we see almost twice the length of that with no drift (and four times the laser pulse). Even at 60 pC the electron beam bunch length is longer than the reference UV laser bunch length of 3.8 ps. The calculated ASTRA-Elegant values at 250 pC/bunch are about 30% lower than those observed in both configurations [10].

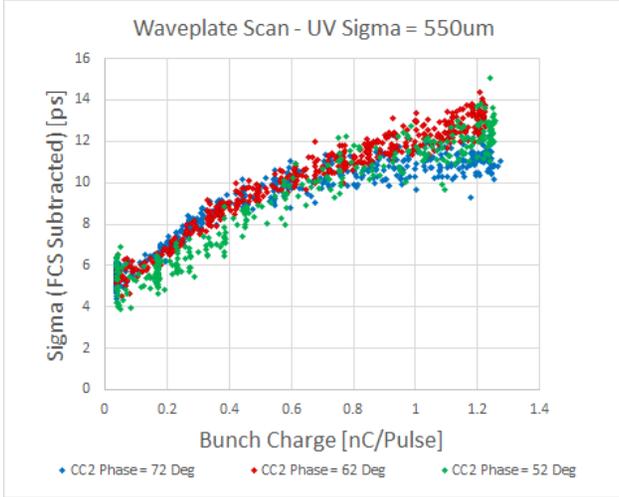

Figure 4: Example of bunch length vs charge at X121 for streak range R2. The UV laser spot was 550 μm (σ).

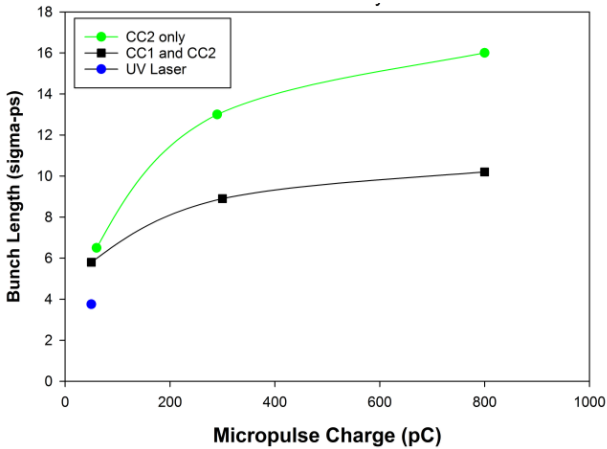

Figure 5: Plot of bunch length vs charge without CC1 and with CC1, or 1.5-m drift and no drift, respectively.

### *First observation of FAST sub-ps bunch lengths*

As part of our studies, we also investigated generation of sub-ps micropulses by transporting through the chicane with rf chirp. As seen in Fig. 6, at 21 degrees off crest and with a micropulse charge of 60 pC, we achieve this target with a synchronous sum of 100 micropulses. During the test we successfully increased the R1 rf amplifier gain to give a new calibration factor of 0.10 ps /pixel and a corresponding camera resolution of ~0.5 ps. A 550-nm longpass filter was used to evaluate and reduce the chromatic temporal dispersion effect at the minimum point. For the data series in the plot, we subtracted out the deduced 1-ps (sigma) bandwidth term in quadrature.

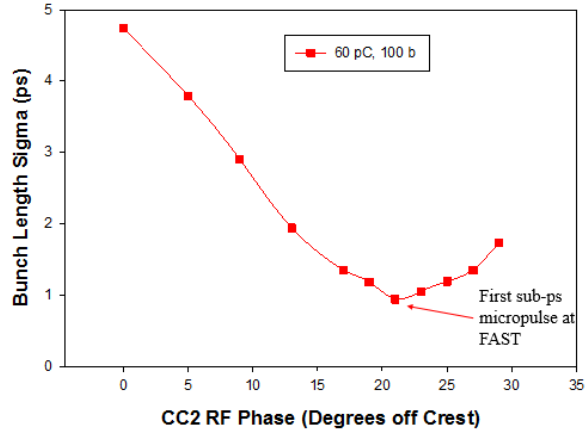

Figure 6: Observation of first FAST sub-ps micropulse bunch length at low charge following chicane compression.

## SUMMARY

In summary, we have performed initial studies of the effects of space-charge-induced electron bunch elongation in the FAST SCRF linac. The bunch elongation was observed to be larger without CC1 installed and with a drift of 1.5 m after the gun. We also observed charge density effects by changing the laser spot size at the cathode. We have initial simulation results that predict a smaller effect for the drift configurations than observed, so further modeling studies are warranted. In addition, we generated a sub-ps micropulse at low charge which also relates to the needed modeling effort. The iteration should lead to a better understanding of the space-charge phenomena.

## ACKNOWLEDGMENTS

The authors acknowledge C. Briegel and J. Diamond for their support for streak camera and readout camera controls, and B. Chase, A. Valishev, D. Broemmilsiek, N. Eddy, R. Dixon, among the many others in the Fermilab AD & TD support departments that made these studies possible. This research is dedicated *in memoriam* to Helen Edwards.